\PassOptionsToPackage{unicode}{hyperref}
\PassOptionsToPackage{hyphens}{url}
\PassOptionsToPackage{dvipsnames,svgnames,x11names}{xcolor}
\documentclass[
]{article}
\usepackage{amsmath,amssymb}
\usepackage{lmodern}
\usepackage{iftex}
\ifPDFTeX
  \usepackage[T1]{fontenc}
  \usepackage[utf8]{inputenc}
  \usepackage{textcomp} 
\else 
  \usepackage{unicode-math}
  \defaultfontfeatures{Scale=MatchLowercase}
  \defaultfontfeatures[\rmfamily]{Ligatures=TeX,Scale=1}
\fi
\IfFileExists{upquote.sty}{\usepackage{upquote}}{}
\IfFileExists{microtype.sty}{
  \usepackage[]{microtype}
  \UseMicrotypeSet[protrusion]{basicmath} 
}{}
\makeatletter
\@ifundefined{KOMAClassName}{
  \IfFileExists{parskip.sty}{%
    \usepackage{parskip}
  }{
    \setlength{\parindent}{0pt}
    \setlength{\parskip}{6pt plus 2pt minus 1pt}}
}{
  \KOMAoptions{parskip=half}}
\makeatother
\usepackage{xcolor}
\NewDocumentCommand\citeproctext{}{}
\NewDocumentCommand\citeproc{mm}{%
  \begingroup\def\citeproctext{#2}\cite{#1}\endgroup}
\makeatletter
 \let\@cite@ofmt\@firstofone
 \def\@biblabel#1{}
 \def\@cite#1#2{{#1\if@tempswa , #2\fi}}
\makeatother
\newlength{\cslhangindent}
\newlength{\csllabelwidth}
\newenvironment{CSLReferences}[2] 
 {\begin{list}{}{%
  \setlength{\itemindent}{0pt}
  \setlength{\leftmargin}{0pt}
  \setlength{\parsep}{0pt}
  \ifodd #1
   \setlength{\leftmargin}{\cslhangindent}
   \setlength{\itemindent}{-1\cslhangindent}
  \fi
  \setlength{\itemsep}{#2\baselineskip}}}
 {\end{list}}
\usepackage{calc}

\ifLuaTeX
\usepackage[bidi=basic]{babel}
\else
\usepackage[bidi=default]{babel}
\fi
\babelprovide[main,import]{american}

\def\languageshorthands#1{}
\ifLuaTeX
  \usepackage{selnolig}  
\fi
\IfFileExists{bookmark.sty}{\usepackage{bookmark}}{\usepackage{hyperref}}
\IfFileExists{xurl.sty}{\usepackage{xurl}}{} 
\hypersetup{
  pdftitle={A Python client for the ATLAS API},
  pdfauthor={Heloise F. Stevance, Jack Leland, Ken W. Smith},
  pdflang={en-US},
  colorlinks=true,
  linkcolor={Maroon},
  filecolor={Maroon},
  citecolor={Blue},
  urlcolor={Blue},
  pdfcreator={LaTeX via pandoc}}

\title{A Python client for the ATLAS API}

\definecolor{c53baa1}{RGB}{83,186,161}
\definecolor{c202826}{RGB}{32,40,38}

\usepackage[affil-it]{authblk}
\usepackage{orcidlink}
\author[1,2%
  *%
  ]{Heloise F. Stevance%
    \,\orcidlink{0000-0002-0504-4323}\,%
    }
\author[3%
  *%
  ]{Jack Leland%
    \,\orcidlink{0000-0001-9262-3587}\,%
    }
\author[1,2%
  \ensuremath\mathparagraph]{Ken W. Smith%
    \,\orcidlink{0000-0001-9535-3199}\,%
    }

\affil[1]{Astrophysics sub-Department, Department of Physics, University
of Oxford, Keble Road, Oxford, OX1 3RH, UK%
  }
\affil[2]{Astrophysics Research Center, Queen's University Belfast,
Belfast, BT7 1NN, UK%
  }
\affil[3]{Oxford Research Software Engineering Group, Doctoral Training
Centre, University of Oxford, Keble Road, Oxford, OX1 3RH, UK%
  }
\affil[$\mathparagraph$]{Corresponding author: %
}
\affil[*]{These authors contributed equally.}
\date{15 May 2025}

\begin{document}
\maketitle

\section{Summary}\label{summary}

The Asteroid Terrestrial-impact Last Alert System (ATLAS) is an all-sky
optical sky survey with a cadence of 24 to 48 hours
(\citeproc{ref-tonry2018}{Tonry et al., 2018}), and the ATLAS Transient
Server (\citeproc{ref-smith2020}{Smith et al., 2020}) processes the
alert stream to enable the discovery and follow-up of extra-galactic
transients. The data from the ATLAS server can be accessed through a
REST API, which has allowed the development of bots that need direct
access to the data to help rank alerts and trigger follow-up
observations of promising targets. Here we present the python client we
have developed for the ATLAS API to help connect bots and scientists to
our data.

\section{Statement of need}\label{statement-of-need}

\texttt{atlasapiclient} is a python client that facilitates the use of
the ATLAS REST API. It provides a class-based interface to all the
read-write utilities of the API and abstracts away the endpoint URLs and
the token management. The \texttt{atlasapiclient} was initially designed
to be used in our transient stream processing pipeline, particularly for
the ATLAS Virtual Research Assistant
(\citeproc{ref-heloise_2025vra}{Heloise \& genghisken, 2025}), but it
can now also be used to connect other astronomy projects to the ATLAS
data and its stream. It is currently allowing the follow-up of ATLAS
alerts by the Mookodi telescope in the South African Astronomical
Observatory (\citeproc{ref-erasmus2024spie}{Nicolas Erasmus et al.,
2024}), which has allowed automated triggering and classification of
transients within 100 Mpc (e.g. (\citeproc{ref-class2025arc}{N. Erasmus
et al., 2025}), (\citeproc{ref-class2025cy}{Wet, 2025})), since early
2025. In the future this will allow us to connect our stream to other
surveys and follow-up facilities (e.g.
(\citeproc{ref-soxs}{Radhakrishnan Santhakumari et al., 2024}))

We expect the API to evolve over time which could break the production
codes that connect to the ATLAS servers. By having a dedicated client
package that includes a full set of unit and integration tests we can
release updates to the client that are compatible with the new API but
do not require users to change their existing code. Decoupling the
user's code form the implementation of the API therefore increases
robustness from the users side.

\section{Data Access}\label{data-access}

In order to gain access to the servers, prospective users will need to
fill a \href{https://forms.gle/Jvy18eejkvxmcN2f6}{Data Request Form}
including a short (no longer than 1 page) science case justifying their
access needs (length of time; Read-only or Read-Write access). We have
also included data policies compliant with the General Data Protection
Regulation (GDPR).

\section{Acknowledgements}\label{acknowledgements}

HFS and JL are supported by the Schmidt Sciences foundation. KS is
supported by the Royal Society.

\section*{References}\label{references}
\addcontentsline{toc}{section}{References}

\phantomsection\label{refs}

\end{document}